\documentclass[a4paper,11pt]{article}
\pdfoutput=1 

\usepackage{jheppub} 

\newcommand{\ba}{\begin{eqnarray}}
\newcommand{\ea}{\end{eqnarray}}

\usepackage[T1]{fontenc} 

\title{\boldmath Gravitational phase transition mediated by thermalon in Einstein-Gauss-Bonnet-Maxwell -Kalb-Ramond gravity}

\author[a,b]{Daris Samart}
\author[b,c,d,e]{Phongpichit Channuie}

\affiliation[a]{Department of Physics, Faculty of Science, Khon Kaen University, Khon Kaen, 40002, Thailand}
\affiliation[b]{School of Science, Walailak University, Thasala, Nakhon Si Thammarat, 80160, Thailand}
\affiliation[c]{College of Graduate Studies, Walailak University, Thasala, Nakhon Si Thammarat, \\80160, Thailand}
\affiliation[d]{Research Group in Applied, Computational and Theoretical Science (ACTS), Walailak University, Thasala, Nakhon Si Thammarat, 80160, Thailand}
\affiliation[e]{Thailand Center of Excellence in Physics, Ministry of Higher Education, Science, Research and Innovation, Bangkok 10400, Thailand}

\emailAdd{darisa@kku.ac.th, channuie@gmail.com}

\abstract{In this work, we  study the possible existence of gravitational phase transition from AdS to dS asymptotic geometries in Einstein-Gauss-Bonnet gravity by adding the Maxwell one-form field ($A_\mu$) and the Kalb-Ramond two-form field ($B_{\mu\nu}$) as impurity substitutions. The phase transitions proceed via the bubble nucleation of spherical thin-shells described by different branches of the solutions which host a dS black hole in the interior and asymptotic thermal AdS state in the exterior. We analyze the phase diagrams of the free energy and temperature to demonstrate the existence of the phase transitions in the grand canonical ensemble (fixed electrical potential). The phase transitions of having the one-form and two-form charges are possible in which the critical temperature is lower than that of the neutral case. Comparing results with existing literature, more importantly, our analyses show that the critical temperature and the Gauss-Bonnet coupling $\lambda$ of the phase transitions get decreased by adding more types of the charges.}

\begin{document} 
\maketitle
\flushbottom

\section{Introduction}
The gravitational phase transition plays a crucial role in both black hole physics and cosmology. On the one hand, the Hawking-Page (HP) phase transition \cite{Hawking:1982dh} is a well known phase transition emerged between the thermal AdS and the AdS black hole. It occurs when temperature reaches its critical value. Then the thermal AdS decays to the black hole and becomes stable. Several types of black hole phase transitions provide rich phenomena which lead to a better understanding of the laws of black hole thermodynamics and provide some hints of the quantum theory of gravity. Moreover, it has been shown that the HP phase transition of the five-dimensional AdS spacetime can be related to the confinement/deconfinement phases in the QCD theory \cite{Witten:1998zw,Nojiri:2001aj}. Therefore, according to the AdS/CFT correspondence paradigm \cite{Maldacena:1997re}, a study of the phase transitions of the higher-dimensional AdS black hole received a huge number of attentions. This is because one might expect to see some interesting effects in the strongly interacting conformal field theory on the boundary (physical real world) from the higher-dimensional gravitational theories in the AdS bulk spacetime. On the other hand, in cosmology, the phase transition is also an very important topic in order to understand the co-existence of different asymptotic vacua \cite{Coleman1977,Coleman1980}. This phase transition may give rise to a possible non-zero vacuum expectation values of the corresponding matter fields (e.g. scalar \cite{Linde:1974at,Veltman:1974au}  and $p$-form \cite{Aurilia:1980xj,Duff:1980qv} fields), contributing to the vacuum energy density or the cosmological constant. There are several mechanisms which were so far proposed to describe how the phase transition happen between different vacua via bubble nucleation processes. This process is driven by the quantum mechanical tunnelling via the instanton \cite{Brown:1987dd,Brown:1988kg}, or it is generated from the thermally triggered transition \cite{Linde:1977mm,Linde:1980tt,Linde:1981zj}.

Thermalon, the Euclidean sector of the thermally activated bubble nucleation \cite{Gomberoff:2003zh,Kim:2007ix,Gupt:2013poa}, is one of compelling mechanisms for mediating gravitational phase transition. More interestingly in higher-order theories of gravity, a number of recent studies have focused on thermalon mediated phase transitions \cite{Cvetic:2001bk,Nojiri:2001pm,Camanho:2012da,Camanho:2015zqa,Camanho:2013uda} in many cases of Lovelock gravity with a vacuum solution. These types of phase transitions proceed through the nucleation of the spherical thin-shell bubbles, so-called thermalon. It is worth noting here that the thermalon, and the techniques associated to it was first introduced in Ref.\cite{Gomberoff:2003zh}. This thin-shell stays between two regions described by different branches of the solutions which host the black hole in the interior. Additionally, the thermalon is a finite temperature instanton which is considered as a thermodynamic phase and described an intermediate state. In a finite time, when the thermalon forms, it is dynamically unstable and then expands to occupy entire space. Hence this effectively changes the asymptotic structure of the spacetime. Once the cosmological constant is fixed, it was shown in Refs.\cite{Camanho:2015zqa,Nojiri:2001pm,Cvetic:2001bk} that thermal AdS space underwent a thermalon-mediated phase transition to an asymptotically dS black hole geometry. This is a so-called the generalize HP (gravitational) phase transition which is different from original HP phase transition for the fact that the thermal AdS vacuum decays into a black hole belonging  to a different (dS) branch solution.

In order to investigate the thermal AdS to dS black hole phase transition as done in Refs.\cite{Camanho:2015zqa,Camanho:2013uda,Camanho:2012da,Hennigar:2015mco} for the neutral model, we shall take a short overview of the mechanisms of the gravitational phase transition in the literature. The initial thermal AdS (outer geometry) will decay and transit to the black hole inside the dS spacetime (inner geometry) via the thermalon mediation. After the thermalon or the bubble (thin-shell) in the Euclidean sector is formed, it will expand eventually reaching the cosmological horizon entirely. At the end, the boundary of a whole spacetime is changed from AdS to dS geometries, i.e., the cosmological constant changes sign from negative to positive values. Therefore, the observer inside the cosmological horizon can measure the thermodynamics quantities of the dS spacetime. One may conclude that the thermalon changes the solutions from one branch to another via the phase transition. More importantly, it has been shown that an reversible process for AdS to dS phase transition does not occur, see more detail discussions in Refs.  \cite{Camanho:2015zqa,Camanho:2013uda,Sierra-Garcia:2017rni,Hennigar:2015mco}. For example, a so-called re-entrant phase transition process in the study of black hole thermodynamics \cite{Altamirano:2013ane,Frassino:2014pha} is not possible. 

So far, a study of gravitational AdS to dS phase transition mediated by thermalon has been done in the neutral case in various aspects \cite{Camanho:2012da,Camanho:2015zqa,Camanho:2013uda,Hennigar:2015mco,Sierra-Garcia:2017rni}. However, these references are considered in the vacuum solution of the Einstein-Gauss-Bonnet gravity. There is a conclusion Ref.\cite{Camanho:2015zqa} that the phase transition of the asymptotically AdS to dS geometries is a generic profile of the higher order gravitational theories without introducing any kind of the matter fields. Nevertheless, it is also speculated that an inclusion of the matter fields in the theory might be useful to embed this model in the string theory. Recently, the phase transition with with the Maxwell field with the fixed charged ensemble is studied in detail Ref.\cite{Samart:2020qya} where the Maxwell or one-form field is the simplest form of the vector gauge field. The results of this study have shown the presence of the Maxwell or static charge decreasing the maximum temperature of the thermalon mediated phase transition and behaving like the impurity substitution in condensed matter physics. However, for the completeness, a study of the grand canonical ensemble of this system is required for investigating the AdS to dS phase transition with the present of the charge. Therefore, it is worth for further investigating the phase transition by adding higher $p$-form fields in the model with the grand canonical ensemble. More importantly, this has not been studied yet in the literature.

The main purpose of this work is to investigate the phase transition profile of this scenario by considering the the Maxwell $U(1)$ one-form field ($A_\mu$) and an additional higher $p$-form field with $U(1)$ gauge group in the grand canonical ensemble. We then choose the Kalb-Ramond (KR) anti-symmetric two-form field ($B_{\mu\nu}$) which is the simplest extension for including the existing fields in the string theory. The KR field is the higher $U(1)$ gauge field generalized the electromagnetic field from particles to strings \cite{Kalb:1974yc}. This is good starting point to make this model embedding and getting closer to the string theory. In addition, studies of the higher rank tensor fields have been made shown its relation with the AdS/CFT conjecture \cite{Germani:2004jf}. Furthermore, string theory reveals the naturalness of higher rank tensor fields in its spectrum \cite{Polchinski:1998rq}.

The content of the paper is organized as follows. In section \ref{s2}, we recap basics of Lovelock gravity with the Maxwell and KR fields. Particularly, we focus on a special case of Lovelock gravity in which the action is reduced to five-dimensional Einstein-Gauss-Bonnet gravity with the Maxwell and the KR fields called EGBMKR that are the starting point for the computations of the present work and construct a junction condition of the proposed model. In this section, we also derive the effective potential of the thermalon EGB gravity and examine the thermalon solutions as well as the stability and dynamics of the thermalon. In section \ref{s3}, we study the gravitational phase transition and the relevant thermodynamic quantities in EGBMKR gravity with KR field. Here we examine how the free energy and temperature depend on the coupling indicating the possibility of thermalon mediated phase transition. We conclude our findings in the last section.

\section{Formalism}\label{s2}

\subsection{The Einstein-Gauss-Bonnet gravity with Maxwell and Kalb-Ramond fields}
In this work, we recall the Einstein-Gauss-Bonnet gravity with the presence of the Maxwell and the KR fields (EGBMKR) at $d=5$ which is reduced from the Lovelock gravity at $K=2$  \cite{Camanho:2012da,Camanho:2013uda,Charmousis:2008kc,Garraffo:2008hu}. The total action of the EGBMKR with its boundary term reads,
\begin{eqnarray}
\mathcal{I} &=& \int_{\mathbb{M}} d^5x\sqrt{-g}\left[ -\,\varepsilon_\Lambda\,\frac{12}{L^2} + R + \frac{\lambda\,L^2}{2}\Big( R^2 - 4\,R_{ab}\,R^{ab} + R_{abcd}\,R^{abcd} \Big) \right] 
\nonumber\\
&& -\, \int_{\mathbb{M}} d^5x\sqrt{-g}\,\frac{1}{4}\left[\mathcal{F}_{ab}\,\mathcal{F}^{ab} 
+ \frac{1}{12}\,\mathcal{H}_{abc}\,\mathcal{H}^{abc}\right]
\nonumber\\
&& -\, \int_{\mathbb{\partial M}} d^{4}x\sqrt{-h}\left[ K +\lambda\,L^2
\left\{ J - 2\left( \mathcal{R}^{AB} -\frac12\,h^{AB}\,\mathcal{R}\right)K_{AB}\right\}\right] \,,
\label{EGBM-action}
\end{eqnarray}
where $J\equiv h^{AB}\,J_{AB}$ is the trace of $J_{AB}$ which is constructed from $K_{AB}$ as \cite{Davis:2002gn}
\begin{eqnarray}
J_{AB} = \frac13\left( 2\,K\,K_{AC}\,K_B^C + K_{CD}\,K^{CD}\,K_{AB} - 2\,K_{AC}\,K^{CD}\,K_{DB} - K^2\,K_{AB}\right) ,
\end{eqnarray}
and $\mathcal{R}_{AB}$ is the Ricci tensor (intrinsic curvature) of the hypersurface, $\Sigma$\,. The spacetime indices of the bulk ($d=5$) and hypersurface ($d=4$) are represented by small and captital Latin alphabets, respectively e.g., $a,\,b,\,c,\,\cdots = 0,\,1,\,2,\,3,\,5$ and $A,\,B,\,C,\,\cdots = 0,\,1,\,2,\,3$\,. More importantly, we note that the coefficients of the Lovelock theory for the Einstein-Gauss-Bonnet gravity case are given by
\begin{eqnarray}
c_0 = \frac{1}{L^2}\,,\qquad c_1 = 1\,,\qquad c_2 = \lambda\,L^2\,.
\end{eqnarray}
Since we have identified the cosmological constant $(\Lambda)$ as
\begin{eqnarray}
\Lambda = \varepsilon_\Lambda\,\frac{6}{L^2}\,,
\end{eqnarray}
where $\varepsilon_\Lambda = \pm\,1$\, is the sign of the bare cosmological constant and we use the $\varepsilon_\Lambda = +\,1$ (de-Sitter) of the bare cosmological constant in this work. In addition, the normalization of the gravitational constant such that $16\pi G_N(d-3)!=1$ is implied as in Refs.\cite{Camanho:2015zqa,Camanho:2013uda,Hennigar:2015mco}.

The standard definition of the Maxwell field in the five-dimensional spherical symmetric spacetime and its equation of motions in the vacuum are given by
\begin{eqnarray}
\mathcal{F}_{ab} &=& \frac12 \left(\partial_a A_b - \partial_b A_a\right),\qquad 
A^{a} = \left( \frac{Q_A}{r^{2}}\,,\,0\,,\,0\,,\,0\,,\,0\right),
\nonumber\\
\partial_a\left(\sqrt{-g}\, \mathcal{F}^{ab} \right) &=& 0\,,\qquad \epsilon^{abcde}\,\partial_a \left(\sqrt{-g}\,\mathcal{F}_{de}\right) =0\,,
\end{eqnarray}
where $A^a$ is the vector potential one-form. The field strength tensor $\mathcal{F}$ in terms of differential form is given by,
\begin{eqnarray}
\mathcal{F} = \frac{Q_A}{r^{3}}\,dt\wedge dr\,,
\label{Maxwell-strength}
\end{eqnarray}
where $Q_A$ is the Maxwell electric one-form charge. 

On the other hand, the KR field and relevant equations of motion are defined by \cite{DeRisi:2007dn,ChiouLahanas:2009cs,Do:2018zac},
\begin{eqnarray}
\mathcal{H}_{abc} 
&=& \frac13 \left(\partial_a \mathcal{B}_{bc} - \partial_b \mathcal{B}_{ca} - \partial_c \mathcal{B}_{ab} \right),
\nonumber\\
\partial_a\left(\sqrt{-g}\, \mathcal{H}^{abc}\right) &=& 0\,,\quad \epsilon^{abcde}\,\partial_a\left(\sqrt{-g}\, \mathcal{H}_{cde}\right) =0\,.
\label{KR_eom}
\end{eqnarray}
We will use the following ansatz for the dual transformation property  \cite{DeRisi:2007dn} of the Kalb-Ramond field in 5-dimension as \cite{ChiouLahanas:2009cs,Do:2018zac,Koivisto:2009sd} 
\begin{eqnarray}
\mathcal{H}^{abc} = \frac12\,\epsilon^{abcde}\,\mathcal{B}_{de}\,.
\label{KR-ansatz}
\end{eqnarray}
According to the ansatz in Eq.(\ref{KR-ansatz}), we find,
\begin{eqnarray}
\mathcal{H}_{abc}\,\mathcal{H}^{abc} = 12\,\mathcal{B}_{ab}\,\mathcal{B}^{ab}\,
\end{eqnarray}
and using its equations of motions in Eq.(\ref{KR_eom}). These give the property of the $\mathcal{B}_{ab}$ and the solution of $B^a$ field as
\begin{eqnarray}
\mathcal{B}_{ab} = \frac12 \left(\partial_a B_b - \partial_b B_a\right),\qquad 
B^{a} = \left( \frac{Q_B}{r^{2}}\,,\,0\,,\,0\,,\,0\,,\,0\right).
\end{eqnarray}
The field strength tensor $\mathcal{B}$ is also written in terms of differential form as the $\mathcal{F}$ field by,
\begin{eqnarray}
\mathcal{B} = \frac{Q_B}{r^{3}}\,dt\wedge dr\,,
\label{KR-strength}
\end{eqnarray}
where $Q_B$ is the KR two-form charge. We note that the KR strength field has the same properties and solution as the Maxwell gauge field under dual transformations and in five-dimensional spacetime.

Next, we continuously construct the spherically symmetric solution of the EGBM theory with the KR filed. The line element of is written in the following form,
\begin{eqnarray}
ds^2 = -\,f(r)\,dt^2 + \frac{dr^2}{f(r)} + r^2\,d\Omega^2_{(\sigma),\,3}\,,
\label{line-element}
\end{eqnarray}
where $d\Omega^2_{(\sigma),\,3}$ is the line element of the $3$-dimensional surface of the constant curvature and it is defined by,
\begin{eqnarray}
d\Omega_{(\sigma),\,3}^2 =
\begin{cases}
d\theta^2 + \sin^2\theta\, d\chi^2 + \sin^2\theta\,\sin^2\chi\,d\phi^2 : \sigma =1\,,
\\
d\theta^2 + d\chi^2 + d\phi^2 : \sigma =0\,,
\\
d\theta^2 + \sinh^2\theta\, d\chi^2 + \sinh^2\theta\,\sinh^2\chi\,d\phi^2 : \sigma =-1\,,
\end{cases}
\end{eqnarray}
Using the explicit forms of the Maxwell and KR strength tensors in Eqs.(\ref{Maxwell-strength}) and (\ref{KR-strength}), the solution of the EGBMKR gravity is written in the simple form by introducing the following polynomial as
\begin{eqnarray}
\Upsilon[g] &=& \sum_{k=0}^{K=2} c_k\,g^k= -\frac{1}{L^2} + g + \lambda\,L^2\,g^2 = \frac{\mathcal{M}}{r^{4}} - \frac{\mathcal{Q}_A^2}{r^{6}} - \frac{\mathcal{Q}_B^2}{r^{6}}\,,
\label{polynomial-sol}
\end{eqnarray}
where the $g(r)$ function is related to the metric tensor function $f(r)$ in Eq.(\ref{line-element}) via,
\begin{eqnarray}
g &\equiv& g(r) = \frac{\sigma - f(r)}{r^2}\,.
\label{g-f-relate}
\end{eqnarray}
The parameters $\mathcal{M}$, $\mathcal{Q}_A$ and $\mathcal{Q}_B$ are related to the black hole ADM mass ($M$) and the electric ($Q_A$) and the KR ($Q_B$) charges via,
\begin{eqnarray}
\mathcal{M} = \frac{M}{8\,\pi}\,,\qquad\quad \mathcal{Q}_X^2 = \frac{Q_X^2}{6}\,,\quad X = A\,,~B\,.
\end{eqnarray}
The detail derivation of the $\Upsilon$ solution can be found in Refs. \cite{Charmousis:2008kc,Garraffo:2008hu,Castro:2013pqa,Chernicoff:2016jsu}.
According to the the polynomial in Eq.(\ref{polynomial-sol}), one finds the solutions of $g$ from the above equation as
\begin{eqnarray}
g_\pm \equiv g_\pm(r) = -\,\frac{1}{2\,\lambda\,L^2}\left( 1 \pm \sqrt{1+ 4\,\lambda\left[ 1
+ L^2\left( \frac{\mathcal{M}_\pm}{r^{4}} - \frac{\mathcal{Q}_{A\pm}^2}{r^{6}} - \frac{\mathcal{Q}_{B\pm}^2}{r^{6}} \right)\right]}\,\right).
\label{g-pm}
\end{eqnarray}
Therefore, the solutions of the line elements for inner and outer manifolds in Eq. (\ref{g-f-relate}) are given by 
\begin{eqnarray}
f_\pm \equiv f_\pm(r) = \sigma + \frac{r^2}{2\,\lambda\,L^2}\left( 1 \pm \sqrt{1+ 4\,\lambda\left[ 1
+ L^2\left( \frac{\mathcal{M}_\pm}{r^{4}} - \frac{\mathcal{Q}_{A\pm}^2+\mathcal{Q}_{B\pm}^2}{r^{6}} \right)\right]}\,\right).
\label{metric-f-pm}
\end{eqnarray}
The above solutions of the EMGB gravity with KR field are similar to the dyonic Einstein-Gauss-Bonnet solutions in string theory which are composed of the electric and magnetic charges \cite{Panahiyan:2018gzq,Dutta:2013dca,Cheng:1993wp,Goldstein:2010aw,Jatkar:1995ut}. In addition, we note that there are two branches of the spherical symmetric solutions of the EGBM theory with KR fields. In addition, the effective cosmological constants of these branch solutions are obtained by setting, $\mathcal{M} = \mathcal{Q}_A = \mathcal{Q}_B=0$ in Eq.(\ref{metric-f-pm}). They read,
\begin{eqnarray}
f_\pm(r) &=& \sigma - \Lambda_\pm^{\rm eff}\,r^2\,,
\nonumber\\
\Lambda_\pm^{\rm eff} &=& -\left(\frac{1 \pm \sqrt{1+ 4\,\lambda}}{2\,\lambda\,L^2}\,\right).
\label{eff-CC}
\end{eqnarray}
More importantly, we find that only the minus branch $f_-(r)$ allows the black hole solution as well as recovering the Einstein theory for $\lambda\to\infty$. In contrast to the plus branch $f_+(r)$, according to Eq.(\ref{eff-CC}),this branch of the solution suffers from the Boulware-Deser (BD) ghost instability and the effective cosmological constant of this branch diverges in the $\lambda\to\infty$ limit which is unphysical. Moreover, the effective cosmological constants of the plus and minus branches in Eq.(\ref{eff-CC}) explicitly give negative and positive values, respectively which mean the $f_+(r)$ and $f_-(r)$ solutions corresponding to AdS and dS spaces. We therefore call $f_+(r)$ and $f_-(r)$ braches as the outer and inner manifold in the latter when we will consider and study the gravitational phase transition between two branches solution of the theory.

\subsection{Junction condition: thermalon dynamics and its stability}
In this section, our main purpose is to study the dynamics of unstable bubble thin shell (thermalon) of the EGBMKR gravity which leads to the AdS to dS gravitational phase transition. We then divide the manifold of the spacetime into two regions  and the timelike surface of the manifold is considered in this work. The manifold is divided as $\mathbb{M} = \mathbb{M}_{-} \cup (\Sigma\times\xi) \cup \mathbb{M}_+$ where $\Sigma$ is the junction hypersurface of two regions of the manifolds and the parameter $\xi \in [0,1]$ is used to interpolate both regions. The $\mathbb{M}_+$ and $\mathbb{M}_-$ are outer and inner regions of the manifolds, respectively. The metric tensor, $f_\pm(r)$ are also used to describe geometries of the outer and inner manifolds which are given by Eq.(\ref{metric-f-pm}). One writes two different line elements of the spacetimes that is used to describe AdS outer ($+$) and dS inner ($-$) spacetime as
\begin{eqnarray}
ds_\pm^2 = -f_\pm(r_\pm)\,dt_\pm^2 + \frac{dr_\pm^2}{f_\pm(r_\pm)} + r^2_\pm\,d\Omega_{(\sigma),\,3}^2\,,
\label{out-in-line}
\end{eqnarray}
again $\pm$ correspond to outer and inner spacetimes respectively. In the latter, we will focus our study in five-dimensional spacetime. Next we are going to construct a manifold $\mathbb{M}$ by matching $\mathbb{M}_\pm$ at their boundaries. The boundary of the hypersurfaces $\partial \mathbb{M}_\pm$ is given by
\begin{eqnarray}
\partial \mathbb{M}_\pm := \Big\{ r_\pm = a | f_\pm > 0\,\Big\}
\end{eqnarray}
with parameterizations of the coordinates
\begin{eqnarray}
r_\pm = a(\tau)\,,\qquad\qquad t_\pm = \widetilde{t}_\pm(\tau)\,,
\end{eqnarray}
where $\tau$ is comoving time of the induced line elements of the hypersurface ($\Sigma$) which takes the same form in both of two manifolds $\mathbb{M}_\pm$ at the boundaries, it reads,
\begin{eqnarray}
ds_\Sigma^2 = -d\tau^2 + a^2(\tau)\,d\Omega_{(\sigma),\,3}^2\,.
\label{surface-line}
\end{eqnarray}
One obtains the following constraint,
\begin{eqnarray}
1 = f_\pm(a)\left(\frac{\partial\, \widetilde{t}_\pm}{\partial \tau}\right)^2 - \frac{1}{f_\pm(a)}\left(\frac{\partial a}{\partial \tau}\right)^2\,.
\end{eqnarray}

It has been shown in detail in Refs \cite{Camanho:2013uda,Camanho:2015ysa} that the continuity of the junction condition across the hypersurface in electro-the vacuum case is written in terms of the canonical momenta, $\pi_{AB}^\pm$ as
\begin{eqnarray}
\pi_{AB}^+ - \pi_{AB}^- = 0\,.
\label{junction-eqn}
\end{eqnarray}
We note that the canonical momentum, $\pi_{AB}$ is derived by varying the gravitational action of the boundary with respect to the induced metric, $h_{ab}$ on the hypersurface, $\Sigma$ i.e. \cite{Camanho:2013uda,Camanho:2015ysa},
\begin{eqnarray}
\delta \mathcal{I}_{\partial\mathbb{M}} = -\int_{\partial\mathbb{M}}d^{4}x\,\pi_{AB}\,\delta h^{AB}\,.
\end{eqnarray}
Refs. \cite{Camanho:2013uda,Camanho:2015ysa} have demonstrated the diagonal components of the $\pi_{ab}^\pm$ have some relation between time and spatial parts via the following constraint,
\begin{eqnarray}
\frac{d}{d\,\tau}\left( a^3\,\pi_{\tau\tau}^\pm\right) = 3\,a^2\,\dot a\,\pi_{\varphi_i\varphi_i}^\pm\,,\qquad \varphi_i = \varphi_1\,,\,\varphi_2\,,\,\varphi_3 = \theta\,,\,\chi\,,\,\phi\,.
\end{eqnarray}
Moreover, the co-moving time component of the $\pi_{ab}^\pm$ can be written in the following compact form \cite{Camanho:2015zqa,Camanho:2013uda,Camanho:2015ysa},
\begin{eqnarray}
\Pi^{\pm} = \pi_{\tau\tau}^\pm &=& \frac{\sqrt{\dot a^2 + f_\pm (r)}}{a}\int_0^1d\xi\,\Upsilon'\Bigg[ \frac{\sigma - \xi^2 f_\pm(a) + (1-\xi^2)\,\dot a^2}{a^2} \Bigg]
\nonumber\\
&=& \int_{\sqrt{H-g_-}}^{\sqrt{H-g_+}} dx\,\Upsilon'\big[ H- x^2\big]\,,
\label{Pi-time-component}
\end{eqnarray}
where $\Upsilon'[x] = d\Upsilon[x]/dx$ and $H = (1 + \dot a^2)/a^2$\,. Furthermore, new variables, $\widetilde{\Pi}$ is defined by $\widetilde{\Pi} = \Pi^+ - \Pi^-$\,. Then the junction conditions of the continuity across hypersurface are given by,
\begin{eqnarray}
\widetilde{\Pi} = 0 = \frac{d\widetilde{\Pi}}{d\tau}\,.
\end{eqnarray}
From now on, we will work on the Euclidean signature, i.e. $t\,\to\,i\,t$ for studying the thermalon which is the Euclidean sector of the spherical bubble thin-shell. This gives $\dot a^2 \,\to\,-\dot a^2$ and $\ddot a \,\to\,-\ddot a$\,.\\
The junction condition of the EGBMKR gravity in Eq. (\ref{junction-eqn}) is implied that
\begin{eqnarray}
\widetilde{\Pi} = \Pi_+ - \Pi_- = 0\,\quad \Longrightarrow \,\quad \Pi_+^2 = \Pi_-^2\,.
\label{junction-EMGB}
\end{eqnarray}
Using the results of the $\Pi_\pm$ in Eq. (\ref{Pi-time-component}) and the metric tensor $f_\pm(a)$ in Eq.(\ref{metric-f-pm}) in the junction condition above, we find
\begin{eqnarray}
\dot a^2 + \frac{a^{6}}{12\, \lambda\,  L^2 }\,\frac{\left(g_+ \left(2 \,g_+\, \lambda\,  L^2+3\right)^2-g_- \left(2\, g_-\, \lambda \, L^2+3\right)^2\right)}{(\mathcal{M}_+ -\mathcal{M}_-)-\left(\mathcal{Q}_+^2-\mathcal{Q}_-^2\right)/a^{d-3} } + 1 = 0\,,
\end{eqnarray}
where we have defined the effective charge, $\mathcal{Q}^2$, as 
\begin{eqnarray}
\mathcal{Q}_\pm^2\equiv \mathcal{Q}_{A,\pm}^2 + \mathcal{Q}_{B,\pm}^2\,.
\end{eqnarray}
Furthermore, we can rewrite the junction condition in terms of kinetic and effective potential energies as
\begin{eqnarray}
\Pi_+^2 = \Pi_-^2\,\quad \Longleftrightarrow \,\quad\, \frac12\,\dot a^2 + V(a) = 0\,.
\end{eqnarray}
Then the effective potential $V(a)$ of the junction condition equation is given by
\begin{eqnarray}
V(a) &=& \frac{a^{6}}{ 24\, \lambda\,  L^2 \Big[(\mathcal{M}_+ -\mathcal{M}_-)-\left(\mathcal{Q}_+^2-\mathcal{Q}_-^2\right)/a^{2} \Big] }
\nonumber\\
&&\qquad\times\,\Bigg[ \left(1+ 4\,\lambda\right) g + 4 \left(2 + g \lambda\,  L^2\right)\left(\frac{\mathcal{M}}{a^{4}} - \frac{\mathcal{Q}^2}{a^{6}}\right)\Bigg]\Bigg|_-^+ + \frac{1}{2}\,,
\label{V-form2}
\end{eqnarray}
where the symbol $\big[ \mathcal{O}\big]\big|_-^+$ is defined by
\begin{eqnarray}
\big[ \mathcal{O}\big]\big|_-^+ \equiv \mathcal{O}_+ - \mathcal{O}_-\,.
\end{eqnarray}
Moreover, we continue to evaluate the derivative of the effective potential, $V'(a)$ and it reads,
\begin{eqnarray}
V'(a) &=& \frac{a^5}{24\, \lambda\,  L^2 \left(\mathcal{M}_+ -\mathcal{M}_- - \left(\mathcal{Q}_+^2-\mathcal{Q}_-^2\right)/a^{2} \right)}
\nonumber\\
&&\quad\times\,\Bigg[6\,(1+ 4\,\lambda)\,g + 12\,\frac{\mathcal{M}}{a^{d-1}} 
+ 2 \left(3 + 6\,\lambda\, L^2 g \right)\frac{\mathcal{Q}^2}{a^{2(d-2)}} \Bigg]\Bigg|_-^+
\nonumber\\
&-& \frac{2\,a^3 \left(\mathcal{Q}_+^2-\mathcal{Q}_-^2\right) }{24\, \lambda\,  L^2 \left(\mathcal{M}_+ -\mathcal{M}_- - \left(\mathcal{Q}_+^2-\mathcal{Q}_-^2\right)/a^{2} \right)^2}
\nonumber\\
&&\quad\times\left[( 1+ 4\,\lambda)\,g + 4 \left(2 + \lambda\,L^2 g \right)\left(\frac{\mathcal{M}}{a^{4}}
- \frac{\mathcal{Q}^2}{a^{6}} \right)\right]\Bigg|_-^+\,.
\label{div-eff-potential}
\end{eqnarray}
In order to derive the effective potential $V(a)$, we have used the following identities to reduce the power of the $g_\pm$ and expressing the linear power of $g_\pm$\,,
\begin{eqnarray}
g_\pm^3 = \frac{g_\pm }{\lambda\,  L^2}\left(\frac{\mathcal{M}_\pm}{a^{4}}-\frac{\mathcal{Q}_\pm^2}{a^{6}} - g_\pm+\frac{1}{L^2} \right),\quad
g_\pm^2 = \frac{1}{\lambda \, L^2}\left(\frac{\mathcal{M}_\pm}{a^{4}}-\frac{\mathcal{Q}_\pm^2}{a^{6}}- g_\pm + \frac{1}{L^2}\right) .
\end{eqnarray}
In addition, the following identities have been used for deriving the derivative of the effective potential, $V'(a)$\,,
\begin{eqnarray}
g' = -\frac{2}{\Upsilon'[g]}\left( \frac{2\,\mathcal{M}}{a^5} - \frac{3\,\mathcal{Q}^2}{a^{7}} \right),
\qquad
\Upsilon'[g] &=& 1 + 2\,\lambda\,L^2\,g\,.
\end{eqnarray}
\begin{figure}[!h]	
	\includegraphics[width=15cm]{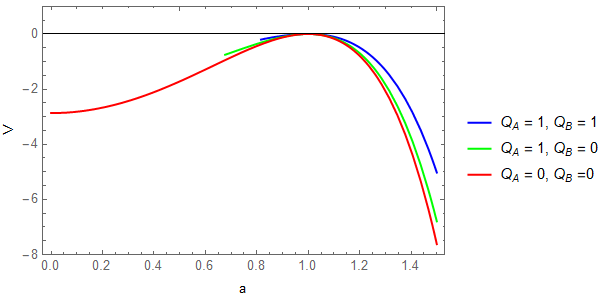}
	\centering
	\caption{The figure displays the shapes of the effective potential of thermalon in values of the charges $\mathcal{Q}_A$ and $\mathcal{Q}_B$ with $\lambda = 0.05$, $a_\star=1$ and $L=1$, $d=5$ and $\sigma=1$.}
	\label{thermalon-potential}
\end{figure}
We note here that the wormhole solution in EGB gravity with Maxwell field has been studied in Ref.\cite{Giribet:2019dmg} considering  the static electric charge ($Q$), i.e., ${Q}_+ = -\,{Q}_-$\,. This condition is hold since the radial direction in one of the asymptotic regions is opposite to the other. However, in our case, there is no charge at the boundary (bubble) and the continuity of both Maxwell ($\mathcal{Q}_{A}$) and KR ($\mathcal{Q}_{B}$) charges acrossing the hypersurface are governed by
\begin{eqnarray}
\mathcal{Q}_A = \mathcal{Q}_{A,+} = \mathcal{Q}_{A,-}\,,\qquad \mathcal{Q}_B = \mathcal{Q}_{B,+} = \mathcal{Q}_{B,-}\,.
\end{eqnarray}
Before we continue to study the stability of the thermalon potential. It is worth to discuss the effect of the presences of the Maxwell and KR fields on the thermalon effective potential displaying in figure \ref{thermalon-potential}. The effective potential of the thermalon with the Maxwell field has been studied in Ref.\cite{Samart:2020qya} and it has been also demonstrated that the inclusion of the static charge $(\mathcal{Q}_A)$ does not change the shape of the effective potential except the existences of the potential. Increasing of the static charge makes the existences of the potential closer to the thermalon position as shown in figure \ref{thermalon-potential}. More importantly, the addition of the KR charge $\mathcal{Q}_B$ further shortens the existence of the effective potential even closer than the presence of the electric (Maxwell) charge $(\mathcal{Q}_A)$ only.

Next we study the stability of the thermalon by considering the linear perturbation (first order) of the effective potential. We note that the thermalon effective potential with $\mathcal{Q}_A$ and $\mathcal{Q}_B$ is similar to the effective potential in Ref.\cite{Samart:2020qya} by replacing $\mathcal{Q}^2\to \mathcal{Q}_A^2 + \mathcal{Q}_B^2$. According to stability analysis of the previous work, the appearances of the charge do not affect the shape of the effective potential of thermalon. 

We close this section with finding the thermalon solutions that useful to study the gravitational phase transition in the latter. We firstly consider the solutions of the thermalon configuration by imposing $V(a_\star) = 0 = V'(a_\star)$\,. Solving those two equations, one obtains the solutions of $\mathcal{M}_\pm$ in terms of $g_\pm^\star$, $a_\star$, $\lambda$, $L$ and $\mathcal{Q}$ as,
\begin{eqnarray}
\mathcal{M}_+(g_\pm^\star,\,a_\star,\,\lambda,\,L^2,\,\mathcal{Q}^2)\; &\equiv& \mathcal{M}_+^\star
\nonumber\\
&=& \frac{1}{4\, \lambda\,L^2\,a_\star^2}
\Big[(1 + 4\,\lambda)\,a_\star^{4}\,\big[  \left(3 + 2\,\lambda\,L^2\,g_-^\star \right)a_\star^{2} 
+ 6 \, \lambda\,L^2 \big]
\nonumber\\
&&\quad\; +\, 4 \left(3 + \lambda\,L^2\, g_-^\star\right)\lambda\,L^2\, \mathcal{Q}^2 + 12\,\lambda ^2\,L^4\, \mathcal{Q}^2 /\, a_\star^{2} \Big],
\label{M_+}\\
\mathcal{M}_-(g_\pm^\star,\,a_\star,\,\lambda,\,L^2,\,\mathcal{Q}^2)\; &\equiv& \mathcal{M}_-^\star
\nonumber\\
&=& \frac{1}{4\, \lambda\,L^2\,a_\star^2}
\Big[(1 + 4\,\lambda)\,a_\star^{4}\,\big[  \left(3 + 2\,\lambda\,L^2\,g_+^\star \right)a_\star^{2} 
+ 6 \, \lambda\,L^2 \big]
\nonumber\\
&&\quad\; +\, 4 \left(3 + \lambda\,L^2\, g_+^\star\right)\lambda\,L^2\, \mathcal{Q}^2 + 12\,\lambda ^2\,L^4\, \mathcal{Q}^2 /\, a_\star^{2} \Big],
\label{M_-}
\end{eqnarray}
Here we used $g_\pm^\star \equiv g_\pm (a_\star)$\,.
Then, we will find the solution of the functions $g_\pm^\star = g_\pm(a_\star)$ in terms of $a_\star$, $\lambda$, $L$, $d$ and $\mathcal{Q}$ via the $\Upsilon[g_\pm] = \mathcal{M}_\pm^\star/a_\star^{4} - \mathcal{Q}^2/a_\star^{6}$ equations. One finds
\begin{eqnarray}
-\frac{1}{L^2} + g_+^\star + \lambda\,L^2\,(g_+^\star)^2 &=& \mathcal{C}_1\,g_-^\star + \mathcal{C}_2\,,
\\
-\frac{1}{L^2} + g_-^\star + \lambda\,L^2\,(g_-^\star)^2 &=& \mathcal{C}_1\,g_+^\star + \mathcal{C}_2 \,,
\end{eqnarray}
where the coefficients $\mathcal{C}_{1,2}$ are given by
\begin{eqnarray}
\mathcal{C}_1 &=& \frac{2\,\lambda\, L^2\, \mathcal{Q}^2 / a_\star^{4}+ a_\star^2\, (1+ 4\,\lambda)}{2\, a_\star^2 }\,,
\\
\mathcal{C}_2 &=& \frac{3\,(1 + 4\,\lambda)\left(a_\star^2 + 2\,\lambda\, L^2\, \sigma \right)+ 4\,\lambda\, L^2\,\mathcal{Q}^2 \left(2\,a_\star^2+ 3\,\lambda\, L^2\, \sigma \right)/a_\star^{6}}{4\,\lambda\, L^2 \,a_\star^2}\,.
\end{eqnarray}
Solving above two equations simultaneously, we obtain the solutions of $g_\pm^\star$ as
\begin{eqnarray}
g_+^\star &=& -\,\frac{(1 + \mathcal{C}_1) + \sqrt{1 + 4\,\lambda - 2\,\mathcal{C}_1 - 3\,\mathcal{C}_1^2 + 4\,\mathcal{C}_2\,\lambda\, L^2}}{2\,\lambda\,L^2}\,,
\label{g+star}
\\
g_-^\star &=& -\,\frac{(1 + \mathcal{C}_1) - \sqrt{1 + 4\,\lambda - 2\,\mathcal{C}_1 - 3\,\mathcal{C}_1^2 + 4\,\mathcal{C}_2\,\lambda\, L^2}}{2 \lambda\,L^4}\,.
\label{g-star}
\end{eqnarray}
We note that $g_-^\star$ has a good behavior (stable) for $\lambda\rightarrow 0$ while $g_+^\star$ gives infinite value (unstable) for $\lambda \rightarrow 0$\,. In addition, we need to study the phase transition between two manifolds of the spacetime, i.e., AdS (outer, $+$) to dS (inner, $-$) then the condition $g_+^\star \neq g_-^\star$ is necessary.

\section{AdS to dS phase transition}\label{s3}
In the present work, there are two different vacua in the model. Specifically, the initial state is thermal anti-de Sitter (AdS) space, while the final one is black hole in de Sitter (dS) spaces. The exterior thermal AdS is initially in the false vacuum state or metastable state and then it decays into black hole inside the interior dS space (true vacuum) via quantum tunneling or jumping across the wall of the quasi particle state in the Euclidean sector called the thermalon. In this work, the decay mechanism proceeds through nucleation of the bubbles or the thermalon of true vacuum (dS) inside the false vacuum (thermal AdS). By using the Lovelock theory of gravity, in particular the Einstein-Gauss-Bonnet gravity, the study of this process has been proposed in Refs.\cite{Camanho:2012da,Camanho:2013uda}, and it was shown that the thermalon effectively jumped from AdS to dS branch solutions with $P \propto e^{-\mathcal{I}_E}$ where $P$ and $\mathcal{I}_E$ being the probability of the decay and the Euclidean action of the difference between initial thermal AdS and the thermalon (bubble state), respectively. This implies that the system with the initial asymptotically AdS geometry will end up in the stable dS black hole after the thermalon expansion filling a whole universe in a finite time and eventually changing asymptotic space to dS geometry with new value of the cosmological constant. The main purpose of this section is to quantify the relevant thermodynamical quantities of the theory and study the possibilities of the AdS to dS gravitational phase transition in the EGBMKR gravity. 

In general, the gravitational phase transition of the charge black hole in the AdS background has been studied in detail by Refs \cite{Chamblin:1999tk,Chamblin:1999hg}. At fixed temperature, this implies that the inclusion of the charge leads us to define the grand canonical ensemble by fixed electric potential, $\Phi$ to describe the thermodynamical system. In the pure thermal AdS, this background consists of charged quanta and it is free to fluctuate during the process. Superficially, this seems to violate the charge conservation however this can be done by fixing electrical potential, $\Phi$ at infinity with respect to the event horizon. We therefore obtain the equation of the grand canonical ensemble to study the thermal properties of this scenario.  

To study the occurrence of the thermalon mediated the AdS to dS phase transition, in the latter, we will consider the free energy of the system in the grand canonical ensemble (fixed electric potential).

\subsection{The grand canonical ensemble (fixed electric potential)}
Before we move forward to calculate the the relevant thermodynamical quantities that use to study the phase transition. We would like to address the event horizon of the dS charge black hole of the interior spacetime. We do this to ensure that the event horizon must locating inside the thermalon radius and the cosmological horizon. More importantly, the radius of the black hole will be used to relate the relevant parameters of the outside and inside spacetimes in order to investigate the phase transitions.  

In the grand canonical ensemble, the charged black hole inside dS geometry is able to create and annihilate charged particles by keeping the electrical potential $\Phi$ fixed until the thermal equilibrium is reached. The electrical potential in $d$-dimension for Maxwell $(\mathcal{Q}_A)$ and KR $(\mathcal{Q}_B)$ charges are given by
\begin{eqnarray} 
\Phi_A = \sqrt{\frac{d-2}{2(d-3)}}\,\frac{\mathcal{Q}_A}{r_B^{d-3}}\,,\qquad 
\Phi_B = \sqrt{\frac{d-2}{2(d-3)}}\,\frac{\mathcal{Q}_B}{r_B^{d-3}}\,,
\label{elec-potential}
\end{eqnarray}
where $r_B$ is black hole radius. To find the black hole radius or the event horizon of the dS black hole $(r_B)$ and the cosmological horizon of the dS spacetime, one solves for $f(r_{H}) = 0$ in the function of the black hole mass inside the interior manifold ($M_-^\star$) as,
\begin{eqnarray}
f_-(r_{H}) = 0\,,\;\Rightarrow \; g_-(r_{H}) = \frac{1}{r_{H}^2} \,,
\end{eqnarray}
where $r_H$ is the radius of the existent horizons of the spacetime. In $d=5$ dimensions, the above equation gives
\begin{eqnarray}
\Upsilon_-\left[ \frac{1}{r_{H}^2}\right] = \frac{\mathcal{M}_-^\star}{r_{H}^{4}} - \frac{\mathcal{Q}^2}{r_{H}^{6}}\,,
\label{solve-rH1}
\end{eqnarray}
where the expression of the $\mathcal{M}_-^\star \equiv \mathcal{M}_-^\star(g_\pm^\star,\,a_\star,\,\lambda,\,L^2,\,\mathcal{Q}^2)$ is given by Eq.(\ref{M_-}) and $\mathcal{Q}^2 \equiv \mathcal{Q}_A^2 + \mathcal{Q}_B^2$. Replacing the charges $\mathcal{Q}_{A,B}$ with the electrical potentials $\Phi_{A,B}$ from Eq.(\ref{elec-potential}) and substituting it into Eq.(\ref{solve-rH1}), the (de-Sitter branch, inner spacetime) horizons, $r_H$ can be obtained from the following equation 
\begin{eqnarray}
r_H^4 - L^2\left(1 + \frac43\, \Phi^2 \right) r_H^2 + L^2\,\big( \mathcal{M}_-^\star - \lambda\,L^2\big)\,r_H^2 = 0\,,
\label{quadratic}
\end{eqnarray}
where the effective electrical potential,$\Phi$ is defined by
\begin{eqnarray}
\Phi^2\equiv \Phi_A^2 + \Phi_B^2\,.
\end{eqnarray} In the present work with the fixed potential case, we limit our study for the non-extremal black hole. We therefore solve the Eq.(\ref{quadratic}) and the radius of the black hole ($r_B$) and cosmological ($r_C$) horizons are given by,
\begin{eqnarray}
r_{B} &=& \frac{L}{\sqrt{6}} \left(\left(4\,\Phi^2 + 3\right) - \sqrt{\left(36\,\lambda 
+\left(4\,\Phi^2 + 3\right)^2\right) - 36\,\frac{\mathcal{M}_-^\star}{L^2}}\,\right)^{\frac12} \,,
\label{rB-g}\\
r_C &=& \frac{L}{\sqrt{6}} \left(\left(4\,\Phi^2 + 3\right) + \sqrt{\left(36\,\lambda 
+\left(4\,\Phi^2 + 3\right)^2\right) - 36\,\frac{\mathcal{M}_-^\star}{L^2}}\,\right)
^{\frac12} \,.
\label{rC-g}
\end{eqnarray}
Note that the solutions of the black hole and cosmological radii have a similar behavior as the neutral case because of Eq.(\ref{quadratic}) takes the same form as the neutral one with additional $\Phi^2$ term. The locations of the event horizon, the thermalon radius and the the cosmological horizon have been demonstrated and confirmed as $r_B < a_\star < r_C$\, \cite{Hennigar:2015mco}. We therefore no need to depict the plot of the mentioned results here.

In addition, one can find the relation between the thermalon radius and black hole in dS interior spacetime as well as other related quantitites in Eqs.(\ref{M_+},\ref{M_-},\ref{g+star},\ref{g-star}) shown earlier. These relations provide us the choice to write down the thermodynamical equations in terms of $a_\star$ and $r_B$ and we will choose to study all quantities as the function of the thermalon radius, $a_\star$.

Having used the on-shell regularization method by subtracting the thermal AdS space background as argued in Refs.\cite{Camanho:2015zqa,Camanho:2013uda,Hennigar:2015mco} for the neutral model in order to calculate the free energy. In the presence of the charges, we keep the electrical potentials $\Phi_{A,B}$ fixed as discussed earlier. This leads to free energy of the thermalon in the grand canonical ensemble with the fixed electrical potential. It reads,
\begin{eqnarray}
F = \mathcal{M}_+ - T_+\,S - \mathcal{Q}_A\,\Phi_A - \mathcal{Q}_B\,\Phi_B\,,
\label{free-energy-grand}
\end{eqnarray}
where $T_+ = 1/\beta_+$ is the Hawking temperature. As shown in above equation, both $\mathcal{M}_+$ and $T_+$ are the quantities corresponding to the exterior observer. Different from the computation of the HP effect in GR, here we have considered the contribution from the thermalon when writing the total Euclidean action \cite{Camanho:2012da,Camanho:2013uda}. This means the thermalon contributes the mass difference between the two branch solutions ($\mathcal{M}_\pm$) but does not contribute to the entropy. In addition, there is no charge on thermolon and the entropy comes from the black hole inside the interior spacetime only. In the following, the free energy of the thermalon is compared to the thermal AdS space where the thermal AdS space is set to zero ($F_{\rm AdS} = 0$) because it was considered to be the background subtraction \cite{Camanho:2015zqa,Camanho:2013uda,Hennigar:2015mco,Sierra-Garcia:2017rni}. By using four conditions, there are two equations $V(a_\star)=0=V'(a_\star)$ from the configurations of the thermalon, Hawking temperature condition to avoid canonical singularity at the horizon, $T = f'(r_B)/4\,\pi$ and the matching temperature of the thermal circle at the thermalon configuration $\beta_+\sqrt{f_+(a_\star)}=\beta_-\sqrt{f_-(a_\star)}$. This leads to only free parameters and we choose $T_+=1/\beta_+$\,. However, the appearances of the Maxwell and KR charges in this work give extra two free parameters i.e. $\mathcal{Q}_A$ and $\mathcal{Q}_B$\,. The Hawking temperature, $T_+$ of the outer space observer is related to the inner space observer by
\begin{eqnarray}
T_+ = \sqrt{\frac{f_+(a_\star)}{f_-(a_\star)}}\,T_-\,,
\label{T_+}
\end{eqnarray}
where the $T_-$ is the Hawking temperature of the inner dS black hole in EMGB gravity. In five-dimensional spacetime and the spherical spatial geometry $\sigma=1$, the $T_-$ is given by \cite{Cai:2003kt},  
\begin{eqnarray}
T_-  =  \frac{1}{4\,\pi\, r_B\sum_{k=0}^{2}k\,c_k\left(\frac{1}{r_B^2}\right)^{k-1}}\,\Bigg[\sum_{k=0}^{2}(4 -2k)\,c_k \left(\frac{1}{r_B^2}\right)^{k-1} - \frac{8}{3}\,\Phi^2\,\Bigg],
\label{T_-}
\end{eqnarray}
We observed that there is a bound of the electrical potential, $\Phi$ which makes the positive values of the temperature $T_-$, i.e.
\begin{eqnarray}
\Phi \leq \sqrt{\frac38\,\left( \sum_{k=0}^{2}(4 -2k)\,c_k \left(\frac{1}{r_B^2}\right)^{k-1} \right)}\,.
\end{eqnarray}
The entropy $S$ is given by \cite{Cai:2003kt},
\begin{eqnarray}
S = 4\,\pi \sum_{k=0}^{2}\frac{k\,c_k}{5-2\,k}\left( \frac{1}{r_B^2}\right)^{k-1}\,.
\label{entropy}
\end{eqnarray}
Note that the entropy of the charged black hole has the same form as the neutral black hole \cite{Camanho:2015zqa,Camanho:2013uda,Hennigar:2015mco}. In addition, the mass parameter $\mathcal{M}_-^\star$ is given by Eq.(\ref{M_+}). Using the black hole radius in Eq.(\ref{rB-g}) and substituted into Eqs.(\ref{M_+},\ref{T_+},\ref{entropy}), we obtain all basic ingredients of the thermodynamics quantities as function of thermalon radius and we are ready to study the thermalon properties and the gravitational phase transitions in the thermodynamics phase space.  
\begin{figure}[!h]	
	\includegraphics[width=10cm]{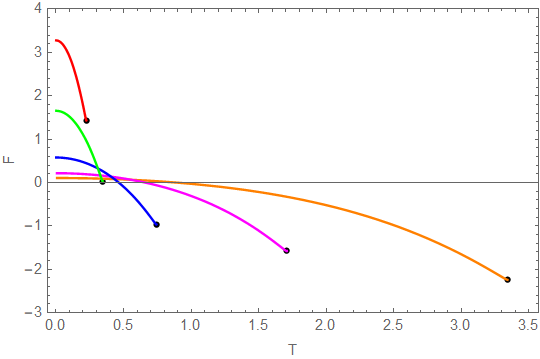}
	\centering
	\caption{The figure displays free energy $F$ of the thermalon configuration as a function of the temperature $T=\beta^{-1}_{+}$ for several values of the coupling $\lambda$. We have used $L=1$ and fixed electircal potential $\Phi \equiv \sqrt{\Phi_A^2 + \Phi_B^2}=0.15$. From right to left: $\lambda=0.05$ (orange), $\lambda=0.10$ (pink), $\lambda=0.25$ (blue), $\lambda=0.65$ (green) and $\lambda=1.20$ (red). The black dots at the end of each plots represent the maximum temperature of the thermalon. All curves correspond to a physical solution $\Pi^+ = \Pi^-$ due to the junction condition of thermalon.}
	\label{FT-grand}
\end{figure}
The free energy is a crucial quantity for investigating the phase transition.  As discussed earlier, we compare the free energy of the thermalon with respect to the free energy of the background subtraction and we set it equal to zero, $F_{\rm AdS} = 0$. This indicates that the sign of the thermalon free energy in Eq.(\ref{free-energy-grand}) corresponding to the phase transition. The plot in figure \ref{FT-grand} shows the free energy of the thermalon configuration as the function of the temperature $T_+$, measured by an exterior observer for several values of the coupling $\lambda$ with the fixed value of the effective electrical potential $\Phi\equiv \sqrt{\Phi_A^2 + \Phi_B^2}$. We have used $L=1$ and $\Phi=0.15$. From right to left: $\lambda=0.05$ (orange), $\lambda=0.10$ (pink), $\lambda=0.25$ (blue), $\lambda=0.65$ (green) and $\lambda=1.20$ (red). All curves in the plot are corresponded to the physical solution at $\Pi^+ = \Pi^-$ of the $V(a_\star)=0=V'(a_\star)$ conditions \cite{Hennigar:2015mco}. The free energy monotonically decreases with increasing temperature when the free energy changes sign from positive to negative implying that the phase transition takes place. We can interpret that the thermalon jumps to the dS branch solution and changes the boundary from AdS to dS asymptotics resulting the discontinuity of the free energy $F$ at the maximum temperature of the physical branch solutions. The behavior of the phase transition occurring in Figure \ref{FT-grand} corresponds to a zeroth-order phase transition \cite{Camanho:2015zqa,Sierra-Garcia:2017rni}. Additionally, we observe that the range of temperatures over which these transitions emerge increases as the coupling $\lambda$ is given smaller with a small charge required. Moreover, thermalon mediated phase transitions are possible over a wide range of temperature for smaller values of the coupling $\lambda$. However, for the given effective electrical potential value $\Phi=0.15$ in figure \ref{FT-grand}, the phase transition is not possible for the coupling $\lambda \gtrsim 0.65$ see green and red lines where the end point of the free energy occur for $F \geq 0$. 

Here we highlight the results in the present work. Due to the inclusion of the additional type of charges, however, the more we add multiple types of the charges, the less possibility we find the phase transition to occur. This phenomenon is analogy to the physical effect in the condensed matter physics as an impurity substitution. For instance, increasing the size of the impurity in a fixed-size host superconductor with a small concentration gives decreasing critical temperature of the host superconductor in the conventional superconductivity \cite{Ghosal:1998,Xiang:1995}. 

\section{Conclusion}
In this work, the toy model of the AdS to dS phase transition in higher-order gravity is considered. It was proposed that the thermalon, the Euclidean spherical bubble thin-shell, changes sign of the effective cosmological constant from the negative value (AdS branch solution) to the positive one (dS branch solution) via the thermal phase transition. This is a so-called the generalization of the Hawking-Page phase transition. Moreover, it is a generic phase transition in higher-order gravity without introducing matter fields. We then investigated the AdS to dS phase transition by adding the Maxwell one-form and the KR two-form fields and analyzed the profile of the phase transition in this framework. Having used the dual transformation in five dimensions, in addition, the KR field can be represented as an additional static charge in this scenario. The junction condition in the model is also constructed. In terms of two static charges, the Maxwell and KR fields do not change the dynamics and stability of the thermalon effective potential except the existences of the effective potential. 

We have studied the phase transitions of the model in the grand canonical ensemble (fixed potential). In this case, we have obtained the relations between free energy ($F$) and temperature ($T_+$) of the grand canonical ensemble revealing the existence of the phase transition with the presence of the charge. Moreover, the maximum temperature and the GB coupling $\lambda$ of the thermalon transition containing two types of charges ($\mathcal{Q}_A$ one-form and $\mathcal{Q}_B$ two-form) are lower than those found in the single charge framework. Interestingly, adding more types of matter fields in higher-order gravity does not change the profile of the phase transition. The inclusion of the charges in the gravitational phase transition is comparable to that of the impurity substitution in condensed matter physics. However we can go beyond a scope of the present work by adding realistic matter fields and considering other modified gravity theories. New features of the gravitational phase transition emerging from those new constitutions can be expected. Intentionally, we leave them for further investigation.

\acknowledgments
This work is financially supported by Walailak University under grant no.WU-CGS-62001.

\end{document}